\documentclass[%
 reprint,
 amsmath,amssymb,
 aps,
]{revtex4-2}

\usepackage{graphicx}
\usepackage{dcolumn}
\usepackage{bm}
\usepackage{float}

\begin{document}

\preprint{APS/123-QED}

\title{Mapping Domain Wall Topology in the Magnetic Weyl Semimetal CeAlSi}

\author{Yue Sun$^{1,2,5}$}
\author{Changmin Lee$^{2}$}
\author{Hung-Yu Yang$^{3}$}
\author{Darius H. Torchinsky$^{4}$}
\author{Fazel Tafti$^{3}$}
\author{Joseph Orenstein$^{1,2}$}
\affiliation{$^{1}$Department of Physics, University of California, Berkeley, California 94720, USA}
\affiliation{$^{2}$Materials Science Division, Lawrence Berkeley National Laboratory, Berkeley, California 94720, USA}
\affiliation{$^{3}$Department of Physics, Boston College, Chestnut Hill, MA 02467, USA}
\affiliation{$^{4}$Department of Physics, Temple University, Philadelphia, Pennsylvania 19122, USA}
\affiliation{$^{5}$Department of Chemistry, University of California, Berkeley, California 94720, USA}
\author{}%
%
%
%
%

\begin{abstract}
We report full vector mapping of local magnetization in CeAlSi, a Weyl semimetal in which both inversion and time-reversal symmetries are broken.  The vector maps reveal unanticipated features both within domains and at their boundaries. Boundaries between domains form two kinds of walls with distinct topology and therefore different interactions with Weyl fermions. Domain walls aligned along the tetragonal axes, e.g. (100), exhibit emergent chirality forbidden by the bulk space group, while diagonal walls are non-chiral.  Within the domains, we observe that the previously reported set of four easy axes aligned along the in-plane diagonals of the tetragonal structure actually split to form an octet with decreasing temperature below the magnetic transition.  All the above phenomena are ultimately traced to the noncollinear magnetic structure of CeAlSi. 

\end{abstract}

\maketitle



Weyl semimetals are condensed matter systems in which nondegenerate bands cross at isolated points, or nodes, in momentum space. Quasiparticles with momenta near the nodes are emergent Weyl fermions, exhibiting linear dispersion and definite chirality \cite{bansil_colloquium_2016,armitage_weyl_2018,wan_topological_2011}.  Although Weyl semimetals generally fall into inversion and time-reversal breaking classes, it is the magnetic, time-reversal breaking class that provides a means to generate and control emergent gauge fields with striking observable consequences \cite{nagaosa_transport_2020,ilan_pseudo-electromagnetic_2020,destraz_magnetism_2020,yuan_discovery_2020}.  

CeAlSi is a hybrid of the two classes of Weyl semimetal introduced above, as its inversion-breaking (tetragonal) crystal structure generates Weyl nodes already in the paramagnetic state. Below the Curie temperature, $T_c$, of $\approx 8.5$ K the local $f$-moments of Ce$^{3+}$ order in a noncollinear ferromagnetic phase. The magnetization, which lies primarily in the ab plane, shifts the momenta of the nodes relative to their positions above $T_c$. Direct evidence for the dependence of nodal momenta on the magnetization direction was seen in the closely related compound CeAlGe, where domain wall resistance gives rise to highly singular structure in the anisotropic magnetoresistance \cite{suzuki_singular_2019}. Recent transport and scanning SQUID magnetometry measurements CeAlSi have reported novel anisotropic anomalous Hall effects \cite{yang_noncollinear_2021} and the existence of magnetic domains with two distinct dynamic magnetic susceptibilities \cite{xu_picoscale_2020}. 

Gauge fields arise in magnetic Weyl semimetals (MWSMs) because the relative separation in momentum space of nodes of opposite chirality is governed by the local magnetization, which acts as an effective vector potential on the chiral charge \cite{liu_chiral_2013,shapourian_viscoelastic_2015,cortijo_elastic_2015,ilan_pseudo-electromagnetic_2020}.  Consequently, considerable attention is focused on magnetic domain walls in MWSMs where temporal and spatial fluctuations of the magnetization are predicted to generate chiral electric and magnetic fields \cite{araki_magnetic_2020,hannukainen_axial_2020,zyuzin_chiral_2015,lux_engineering_2018,tchoumakov_magnetic_2017,liang_topological_2020}. 

Here we report the topology of domain walls in CeAlSi, observed by mapping the magnetization vector field, $\boldsymbol{M}(\boldsymbol{r})$, using a scanning Kerr effect microscope. The magneto-optical Kerr effect (MOKE) is the rotation of the plane of polarization on reflection from a medium with broken time-reversal symmetry \cite{mccord_progress_2015}. At normal incidence, the MOKE signal, $\Theta$, is sensitive only to the out-of-plane ($z$) component of the magnetization.  However, upon changing the beam path to oblique incidence the polarization rotation becomes sensitive to the in-plane components of the magnetization as well \cite{qiu_surface_2000,stupakiewicz_direct_2014,rave_quantitative_1987,yang_combined_1993,daboo_vectorial_1993,ding_experimental_2000}. When all three components of $\boldsymbol{M}$ are present, $\Theta$ is a superposition of the polar, longitudinal, and transverse Kerr effects.  For the measurements reported here, we developed a vector MOKE (VMOKE) method to disentangle these effects and obtain maps of all three components of the local magnetization vector. 
\begin{figure}[t]
\centering
\includegraphics[width=\linewidth]{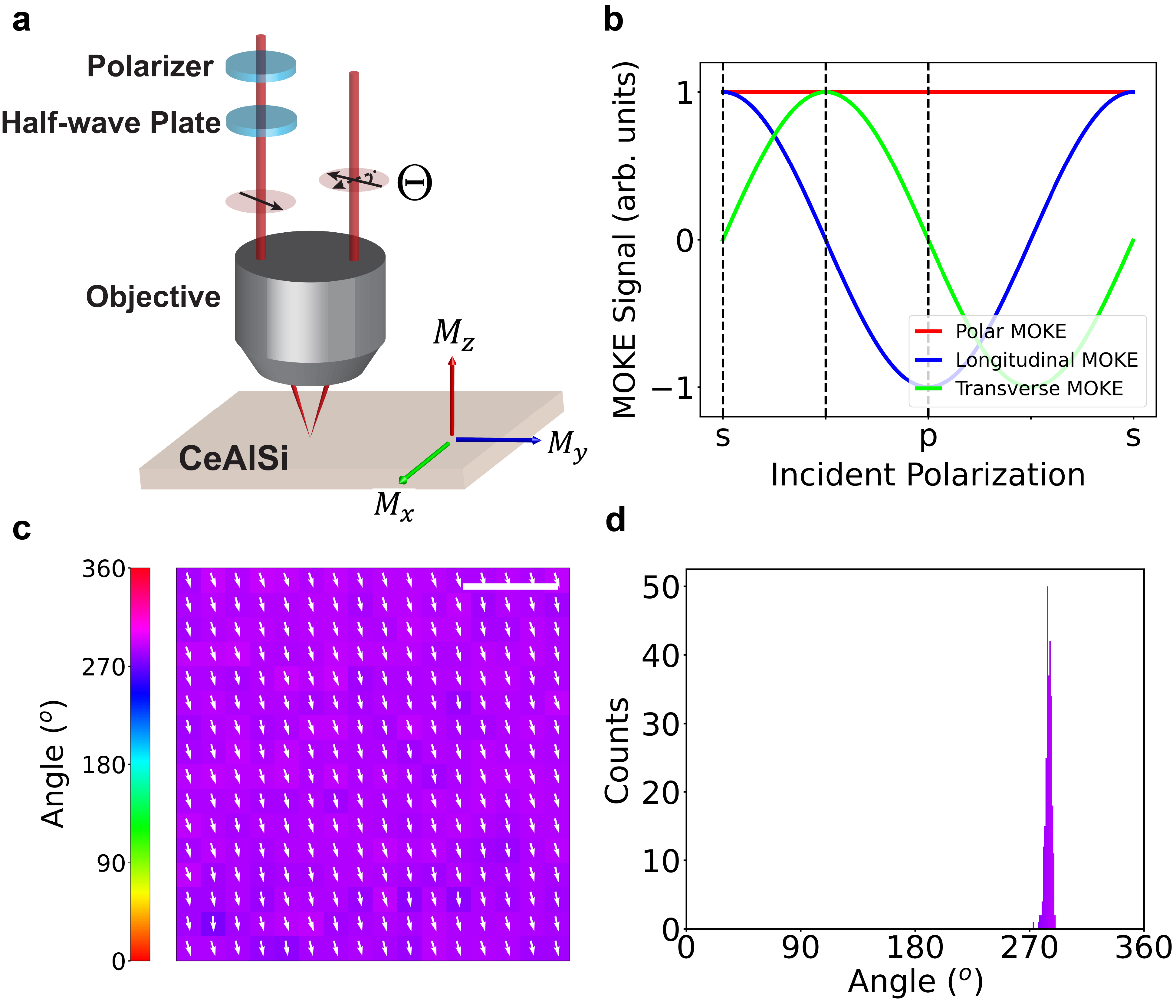}
\caption{\label{fig:epsart} (a) Schematic of the optical set-up of  vector MOKE. The combination of a polarizer and a half-wave plate controls the incident polarization. The incident beam is focused by a 10x microscope objective. (b) Polarization dependence of polar, longitudinal and transverse contributions to the Kerr rotation, $\Theta$. (c) Vector MOKE map of a single domain state in CeAlSi prepared by a 150 Oe magnetic field. The area of the region is 80 $\mu$m$\times$80 $\mu$m and is sampled in 5 $\mu$m steps. Scale bar, 20 $\mu$m. (d) Histogram of the magnetization direction in (c).}
\end{figure}

VMOKE is based on measuring the dependence of $\Theta$ on the plane of linear polarization of the incident light.  Figure 1(a) shows a schematic of the optical set-up. The incident polarization is controlled by a combination of a polarizer and a half-wave plate and $\Theta$ is measured with a balanced optical bridge detector (see Supplementary Information).  Figure 1(b) shows a summary of the polarization dependence of $\Theta$ for the three Cartesian components of $\boldsymbol{M}$, where $yz$ is the plane of incidence.  In the usual convention, $s$ and $p$ polarization denote incident light polarization perpendicular and parallel to the plane of incidence, respectively.  The polarization rotation resulting from $M_{z}$ is independent of the incident polarization.  $M_{y}$, which lies in the plane of incidence, generates a Kerr rotation that switches sign for $s$- and $p$-polarized input beams \cite{rave_quantitative_1987,yang_combined_1993}. Finally, $M_{x}$ generates optical birefringence, leading to rotation on reflection that reverses sign when the incident polarization is rotated by $\pm$45° with respect to the plane of incidence. Based on their distinct polarization dependences, we can determine the three components of $\boldsymbol{M}$ at each location in the sample by performing three measurements: $\Theta(0)$, $\Theta(\pi/4)$, and $\Theta(\pi/2)$ (see Supplementary Information).

To eliminate long-term drifts and enhance sensitivity, we modulate $\Theta$ by overlapping the 780 nm probe beam with a 1560 nm pump beam chopped at 2.5 kHz. Lock-in detection at the chopping frequency allows for measurement of $\Theta$ at the microradian level. In order to validate our VMOKE technique we prepared a single domain state in CeAlSi by applying a 150 Oe magnetic field, which is stronger than the coercive field (70 Oe) \cite{yang_noncollinear_2021}. The direction of the magnetic field (measured by a Hall effect magnetometer) was $\approx$10° from the [010] direction of the sample. The arrows in the vector field map (Figure 1(c)) illustrate the local magnetization direction as determined by scanning VMOKE. Figure 1(d) shows a histogram of the distribution of magnetization direction, showing a narrow peak 280°, which matches the direction of the external magnetic field.

Figure 2(a) presents spontaneous magnetization maps of CeAlSi at different temperatures. The sample was cooled under zero external magnetic field and the maps were measured during warming.  The color code illustrates the direction of the in-plane magnetization $\boldsymbol{M}_{\parallel}$.  The maximum out-of-plane component is approximately 1\% of the in-plane components and will be discussed later.  Clearly evident are large domains, of order 50 $\mu$m across, consistent with measurements performed at 6 K and above by scanning SQUID microscopy, which detects the near surface local magnetic flux \cite{xu_picoscale_2020}. In the maps taken below 5 K, long-distance vertical and diagonal domain walls are observed.  The domain structure changes with temperature variation as small as 0.5 K, and non-monotonically through the temperature range from 5-7 K, as on warming the domain pattern becomes first more disordered and then less disordered.  The large fluctuations suggest that the domain walls are highly mobile at these intermediate temperatures.  The map-average magnetization amplitude $\left|\boldsymbol{M}\right|$ goes to zero at $\approx$ 9 K (Figure 4(b)), which is close to the value of $T_{c}$ (8.2 K) extracted from heat capacity measurement \cite{yang_noncollinear_2021}.
\begin{figure}[t]
\centering
\includegraphics[width=\linewidth]{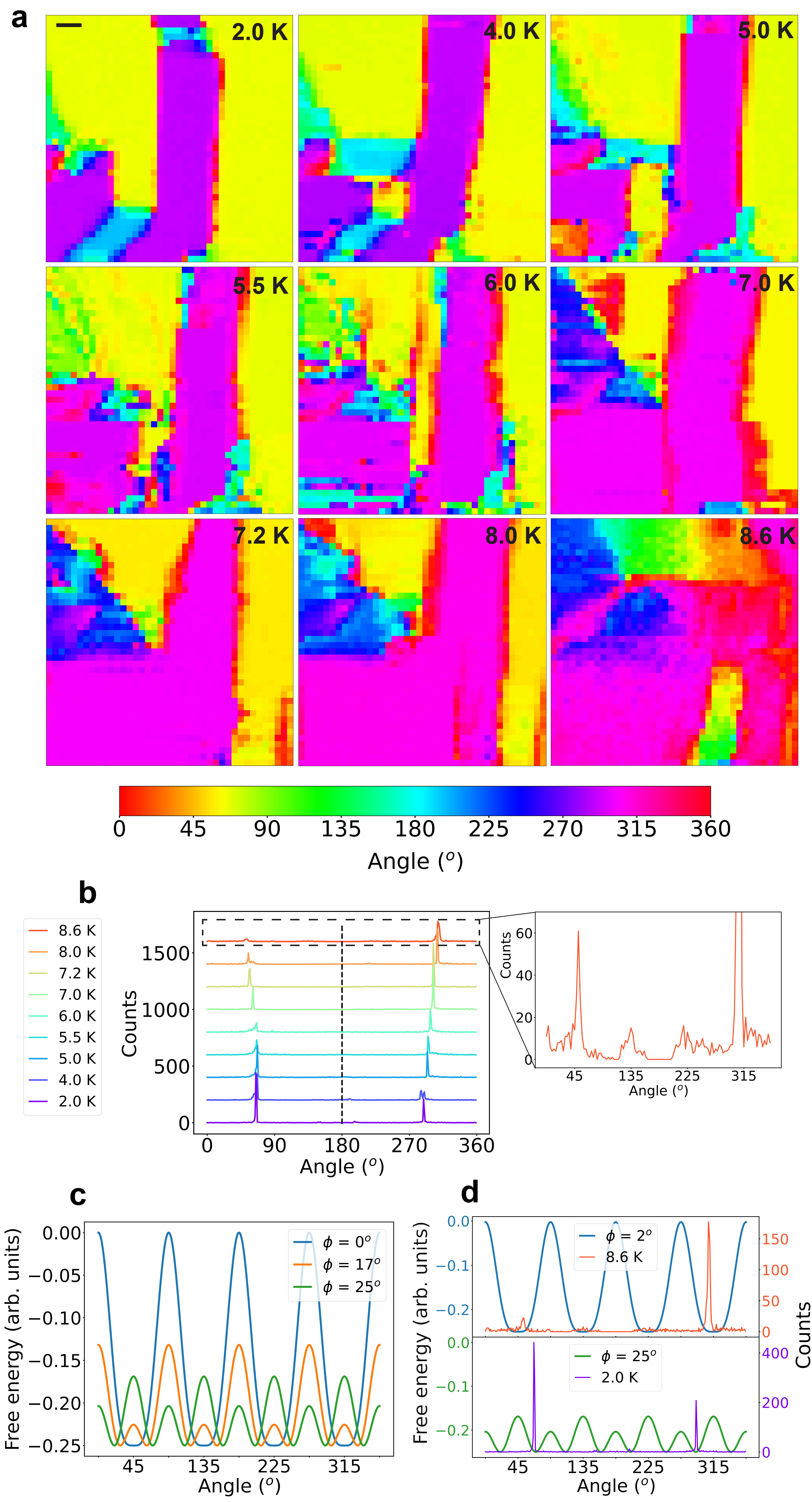}
\caption{\label{fig:epsart} (a) Spontaneous magnetization map across a 200 $\mu$m$\times$200 $\mu$m area of the sample from 2 to 8.6 K. Color code illustrates the direction of the in-plane magnetization. Scale bar, 20 $\mu$m.  (b) Histograms of the distribution of magnetization directions from 2 to 8.6 K. Expanded view, the histogram at 8.6 K. (c) Free energy as a function of $\theta$ with $\gamma=1$, $\phi=$ 0°, 17° and 25°. (d) Comparison between the distribution of the magnetization direction and the free energy at 2 K and 8.6 K.}
\end{figure}

To organize the large information content of the maps of $\boldsymbol{M}(\boldsymbol{r})$, we consider first the orientation of $\boldsymbol{M}$ within the domains.  Previous studies of CeAlSi and the related compound CeAlGe concluded that the in-plane magnetization in zero field is oriented along four easy axis directions: (110) and the other three directions generated by the four-fold symmetry of the $C_{4v}$ point group \cite{yang_noncollinear_2021,suzuki_singular_2019}.  However, the $\boldsymbol{M}(\boldsymbol{r})$ maps reveal that this is not the full story. 

In Figure 2(b), we plot the distribution of magnetization directions at each of the measured temperatures. Two dominant angles are observed at every temperature, reflecting the fact that the maps are dominated by the purple and yellow domains. Some small cyan and green domains are also present in the maps.  At temperatures near $T_{c}$, the dominant angles are close to (110) and symmetry related directions, as can be seen in the expanded view of the 8.6 K data. However, it is clear that the peaks of the distribution begin to depart from these angles with cooling below about 5 K.  At 2.0 K, the histogram peaks at 66° (yellow), 290° (purple), 200° (cyan) and 155° (green), have shifted from the previously reported easy axis directions by  $\approx$25°. As this rotation of easy axes with decreasing temperature takes place, they remain symmetric with respect to 180°, which is labeled as a dashed line in Figure 2(b). 

To understand rotation of the easy axes we construct a Landau free energy model based on the noncollinear magnetic order observed in neutron scattering measurements \cite{yang_noncollinear_2021}. The CeAlSi structure is comprised of two alternating layers of Ce atoms, whose magnetizations are $\boldsymbol{M_{1}}$ and $\boldsymbol{M_{2}}$ separately.  The noncollinear order was shown to derive from an anisotropic g-tensor in isostructural CeAlGe \cite{suzuki_singular_2019}. Since the magnetization is mostly in-plane, we take only the $x$ and $y$ components into consideration. With the inclusion of interlayer coupling, the anisotropy energy $F_{a}$ can be written as the form below, which respects the four-fold rotational symmetry of the structure,
\begin{equation}
\resizebox{0.9\hsize}{!}{$%
F_{a}=-\alpha \left(M_{1x}^{2}M_{1y}^{2}+M_{2x}^{2}M_{2y}^{2}\right)+4\beta M_{1x}^{2}M_{1y}^{2}M_{2x}^{2}M_{2y}^{2}.
$%
}%
\end{equation}
The first term is the anisotropy energy in each layer.  Restricting the free energy to this term, which treats the two layers as independent, yields easy axes parallel to the [110] and symmetry related directions. 

The second term is the interlayer coupling. Assuming ferromagnetic order in each layer, the directions of $\boldsymbol{M_{1}}$ and $\boldsymbol{M_{2}}$ can be expressed as $\theta+\phi$ and $\theta-\phi$, where $\theta$ is the direction of the net magnetization and $2\phi$ is the angle spanned by $\boldsymbol{M_{1}}$ and $\boldsymbol{M_{2}}$. Substituting the two angles into Equation (1), one finds
\begin{equation}
	F_{a}\propto (1-\gamma/2)\cos 4\phi \cos 4\theta +(\gamma/8)(\cos 8\theta+\cos 8\phi),
\end{equation}
where $\gamma\equiv\beta/\alpha$. Figure 2(c) illustrates the free energy as a function of $\theta$ for $\gamma=1$ and values of $\phi$ that are consistent with the neutron scattering measurements. For collinear magnetization ($\phi=0$) the free energy minima occur at [110] and related directions.  The four minima split into eight with increasing $\phi$; note that the eight easy axes continue to respect the rotational and mirror symmetries of the crystal. Figure 2(d) compares the distribution of magnetization angles with the free energy at 2.0 K and 8.6 K, showing that temperature-dependent interlayer coupling captures the rotation of dominant angles observed in the Kerr maps. As a bonus, the model also accounts for the order $\rightarrow$ disorder $\rightarrow$ order transition as the temperature crosses 6 K. We associate the disordered domain patterns with the flatness of the free energy that occurs at the transition from four to eight minima. 
\begin{figure}[t]
\centering
\includegraphics[width=\linewidth]{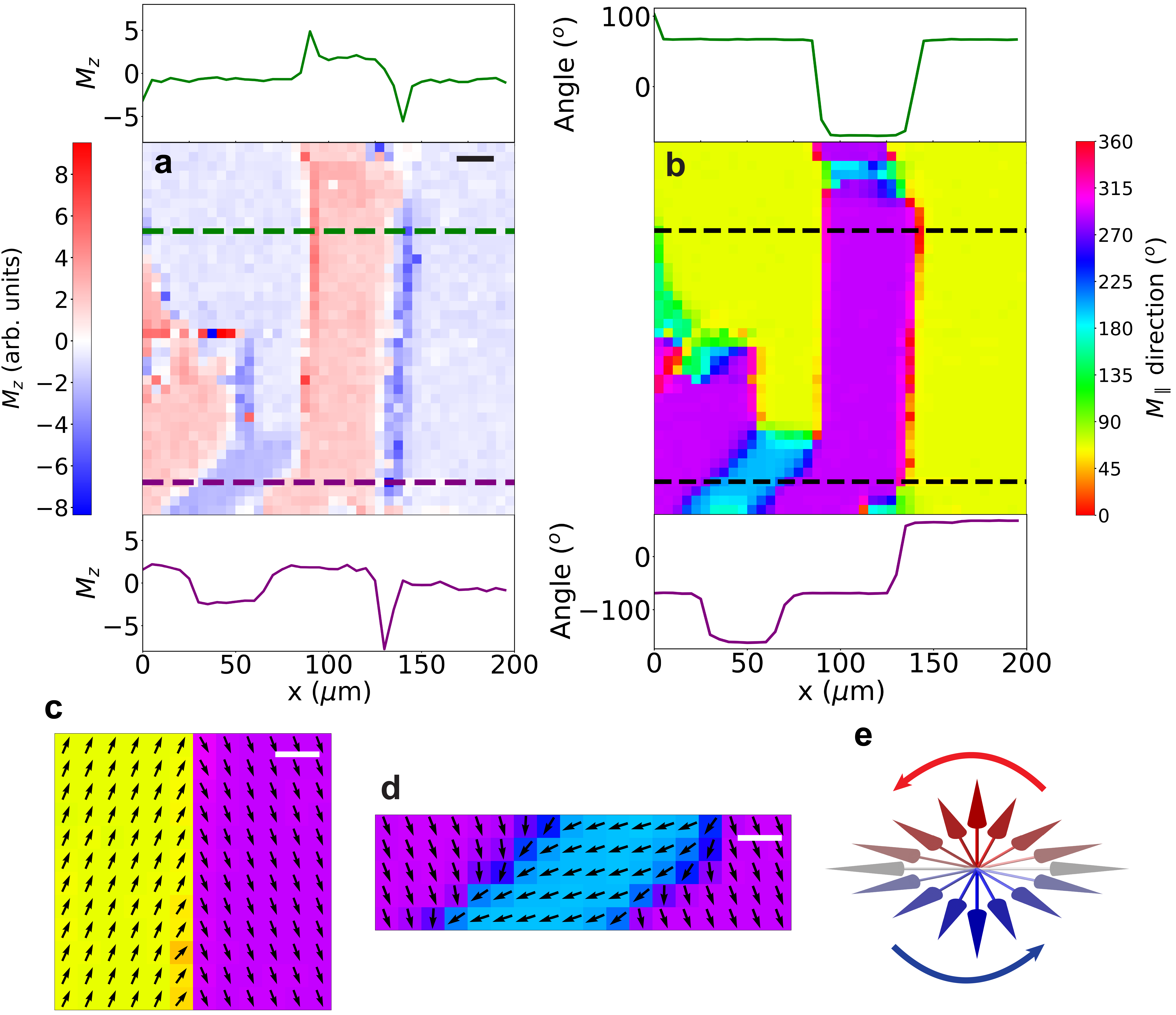}
\caption{\label{fig:epsart} (a) Map of the amplitude of $M_{z}$ component at 2 K. The map is plotting the measured Kerr rotation as the unit of microradian, which is proportional to $M_{z}$. Scale bar, 20 $\mu$m. The line cuts above and below plot $M_{z}$ along the lines with corresponding colors. (b) Spontaneous in-plane magnetization map at 2 K. The line cuts above and below plot the direction of in-plane magnetization along the same lines in (a). (c)(d) Expanded image of vertical and diagonal domain walls respectively. Arrows indicate the direction of in-plane magnetization. (e) Side view of magnetization as the line cut traverses the two vertical domain walls  (with the $z$ component increased for clarity).}
\end{figure}

With this understanding of the magnetization within the domains, we next focus on the variations in $\boldsymbol{M}(\boldsymbol{r})$ that occur at the domain boundaries. Figure 3(a) is a map of $M_{z}$ at 2 K, while Figure 3(b) shows the orientation of the in-plane magnetization measured at the same temperature using the same color scale as in Figure 1(a). The maps reveal vertically and diagonally oriented domain walls, i.e. parallel to [100] and [110], respectively, whose magnetization texture is topologically distinct. The contrasting texture can be seen in the expanded images shown in Figs. 3(c) and 3(d), in which arrows represent the local magnetization direction. 

The contrasting character of the two walls is revealed by comparing the line cuts above and below the map in Figure 3(a). The upper line cut, which traverses two vertical domain walls, shows peaks in $M_{z}$ whose sign depends on the sign of $\partial M_{para}/\partial n$, where $M_{para}$ is the component of magnetization parallel to the wall and $n$ is the normal coordinate. The cartoon in Figure 3(e) shows a side view of magnetization as the line cut traverses the two vertical domain walls  (with the $z$ component increased for clarity). The magnetization vector traces a highly eccentric ellipse when viewed from the plane of the map.  The sense of the rotation is the same for the two boundaries, indicating that the domain walls are chiral. The same sense of chirality was observed for all vertical domains over multiple cool downs. In contrast, the line cuts through the diagonal domain walls depicted below Figure 3(a) show that in this case $M_{z}$ goes continuously through zero. We note that the walls differ as well in their magnetic charge density, $\sigma_M\equiv\boldsymbol{n}\cdot\left(\boldsymbol{M}_{a}-\boldsymbol{M}_{b}\right)$, which is of order $M_\parallel$ in the diagonal walls and zero for the vertical walls, where $\boldsymbol{M}_{a}$ and $\boldsymbol{M}_{b}$ denote the magnetization on both sides of the domain wall. Thus chiral magnetic charge neutral and non-chiral charged domain walls coexist in CeAlSi. 
\begin{figure}
\centering
\includegraphics[width=\linewidth]{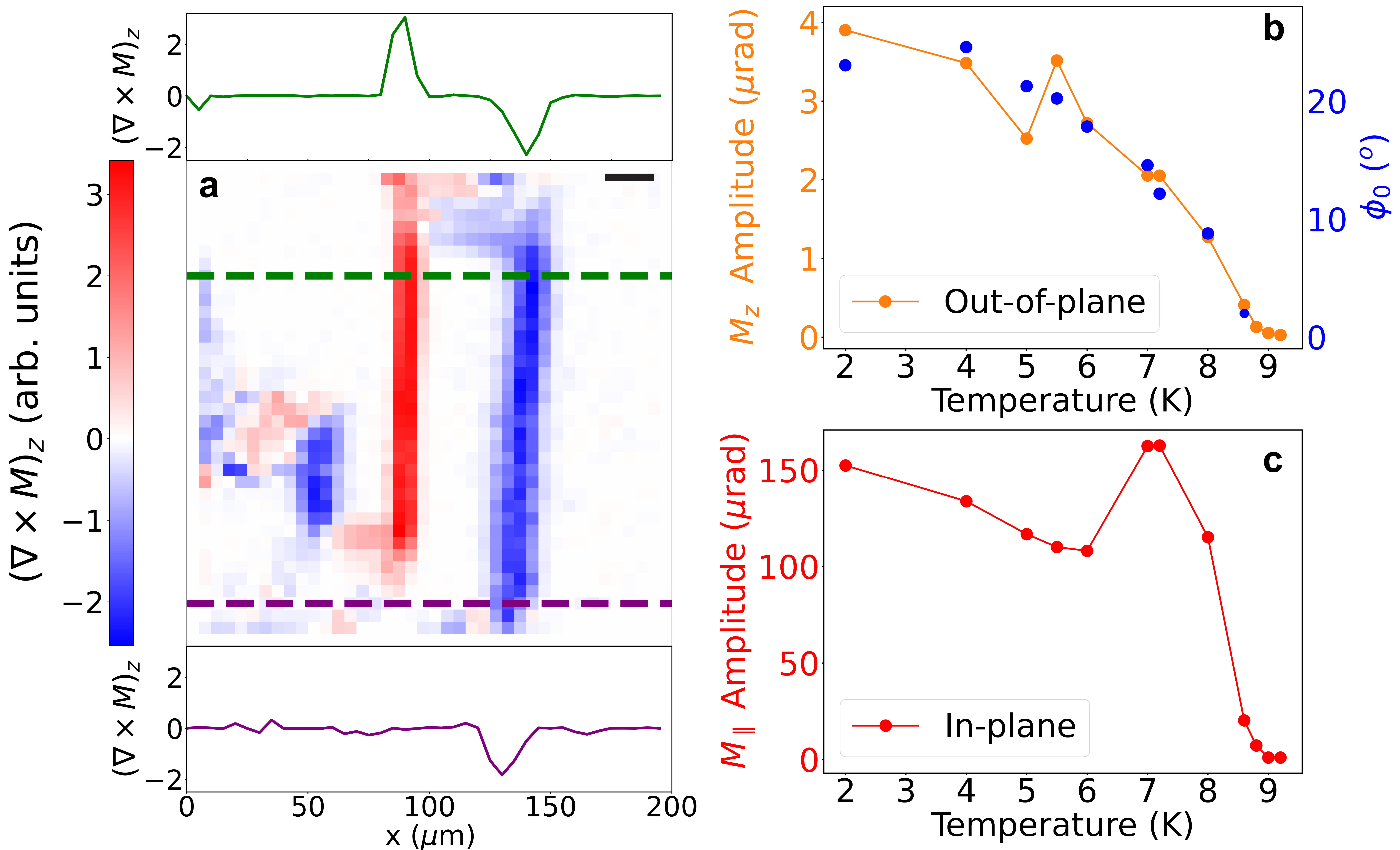}
\caption{\label{fig:epsart} (a) Map of $\left(\nabla\times\boldsymbol{M}\right)_{z}$ at 2 K. Scale bar, 20 $\mu$m. The line cuts above and below plots $\left(\nabla\times\boldsymbol{M}\right)_{z}$ along the lines with corresponding colors. (b)  The amplitude of $M_{z}$ and $\phi$ as a function of temperature. Orange dots: $M_{z}$ amplitude in chiral domain boundaries. Blue dots: $\phi$ extracted by fitting the histogram in each temperature. (c) The map-averaged amplitude of $M_{\parallel}$ as a function of temperature.}
\end{figure}

The peaks in $M_z$ at the vertical domain walls suggest the existence of a local Lifshitz invariant \cite{ullah_crystal_2019} of the form, $DM_{z}\left(\nabla\times\boldsymbol{M}\right)_{z}$, which will induce chiral walls when $\left(\nabla\times\boldsymbol{M}\right)_{z}\neq 0$. We note that such a gyromagnetic term in the free energy is forbidden in the bulk of the crystal by the mirror symmetry that takes $x\rightarrow -x$.  However, the noncollinear ordering of the magnetic bilayers within the unit cell can break the mirror symmetry locally, permitting $DM_{z}\left(\nabla\times\boldsymbol{M}\right)_{z}\neq 0$ (see Supplementary Information). Figure 4(a), which is a map of $\left(\nabla\times\boldsymbol{M}\right)_{z}$, shows that the difference in chirality of the two types of walls is consistent with the picture of a local gyromagnetic invariant. As seen in the line cut through the vertical walls, $\left(\nabla\times\boldsymbol{M}\right)_{z}$ peaks at the domain boundaries, reproducing the structure in $M_z$ shown previously, while the line cut through the  diagonal domain wall shows that $\left(\nabla\times\boldsymbol{M}\right)_{z}$ at the nonchiral boundary.

We have argued above that the existence of a local chirality-generating term in the free energy is a consequence of the noncollinear magnetic ordering. Further support for this hypothesis is seen in Figure 4(b), which compares the temperature dependence of the wall-centered peak in $M_z$ with $\phi$, the angle between the magnetization in adjacent layers as deduced from the shift of the in-plane easy axes. The proportionality of these observables supports a causal relation between in-plane noncolinearity and domain wall chirality. Both quantities onset more gradually with decreasing temperature than the magnetization itself, whose $T$ dependence is shown in Figure 4(c). 

In summary, we have reported full vector imaging of magnetization in the magnetic Weyl semimetal CeAlSi, revealing new properties causally connected to its noncollinear magnetic structure.  Coupling between adjacent noncollinear layers of Ce moments splits the conventional four-fold pattern of in-plane easy axes to an octet and leads to the formation of two classes of domain walls.  The walls exhibit contrasting behavior in both chirality and local magnetic charge density.  In the charge-neutral walls aligned along the in-plane tetragonal crystal axes, the magnetization traces an elliptical orbit as the wall is traversed, while the charged walls that form parallel to (110) are nonchiral.  The existence of walls with distinct topology will enable future tests of the role of magnetic texture in determining emergent gauge fields in Weyl magnets and their coupling to “real” external fields.  Strong hints of distinct responses to external fields corresponding to the two classes of domains walls have already been seen in local measurements of ac susceptibility \cite{xu_picoscale_2020} and serve as additional motivation for future studies.

\begin{acknowledgments}
Optical measurements and analysis were performed at the Lawrence Berkeley Laboratory as part of the Quantum Materials program, Director, Office of Science, Office of Basic Energy Sciences, Materials Sciences and Engineering Division, of the U.S. Department of Energy under Contract No. DE-AC02-05CH11231. J.O and Y.S received support from the Moore Foundation's EPiQS Initiative through Grant GBMF4537 to J.O. at UC Berkeley. The work at Boston College was funded by the National Science Foundation under award No. DMR-1708929. 
\end{acknowledgments}

\bibliographystyle{apsrev4-2}
\bibliography{CeAlSiReference}

\begin{thebibliography}{27}%
\makeatletter
\providecommand \@ifxundefined [1]{%
 \@ifx{#1\undefined}
}%
\providecommand \@ifnum [1]{%
 \ifnum #1\expandafter \@firstoftwo
 \else \expandafter \@secondoftwo
 \fi
}%
\providecommand \@ifx [1]{%
 \ifx #1\expandafter \@firstoftwo
 \else \expandafter \@secondoftwo
 \fi
}%
\providecommand \natexlab [1]{#1}%
\providecommand \enquote  [1]{``#1''}%
\providecommand \bibnamefont  [1]{#1}%
\providecommand \bibfnamefont [1]{#1}%
\providecommand \citenamefont [1]{#1}%
\providecommand \href@noop [0]{\@secondoftwo}%
\providecommand \href [0]{\begingroup \@sanitize@url \@href}%
\providecommand \@href[1]{\@@startlink{#1}\@@href}%
\providecommand \@@href[1]{\endgroup#1\@@endlink}%
\providecommand \@sanitize@url [0]{\catcode `\\12\catcode `\$12\catcode
  `\&12\catcode `\#12\catcode `\^12\catcode `\_12\catcode `\%12\relax}%
\providecommand \@@startlink[1]{}%
\providecommand \@@endlink[0]{}%
\providecommand \url  [0]{\begingroup\@sanitize@url \@url }%
\providecommand \@url [1]{\endgroup\@href {#1}{\urlprefix }}%
\providecommand \urlprefix  [0]{URL }%
\providecommand \Eprint [0]{\href }%
\providecommand \doibase [0]{https://doi.org/}%
\providecommand \selectlanguage [0]{\@gobble}%
\providecommand \bibinfo  [0]{\@secondoftwo}%
\providecommand \bibfield  [0]{\@secondoftwo}%
\providecommand \translation [1]{[#1]}%
\providecommand \BibitemOpen [0]{}%
\providecommand \bibitemStop [0]{}%
\providecommand \bibitemNoStop [0]{.\EOS\space}%
\providecommand \EOS [0]{\spacefactor3000\relax}%
\providecommand \BibitemShut  [1]{\csname bibitem#1\endcsname}%
\let\auto@bib@innerbib\@empty
\bibitem [{\citenamefont {Bansil}\ \emph {et~al.}(2016)\citenamefont {Bansil},
  \citenamefont {Lin},\ and\ \citenamefont {Das}}]{bansil_colloquium_2016}%
  \BibitemOpen
  \bibfield  {author} {\bibinfo {author} {\bibfnamefont {A.}~\bibnamefont
  {Bansil}}, \bibinfo {author} {\bibfnamefont {H.}~\bibnamefont {Lin}},\ and\
  \bibinfo {author} {\bibfnamefont {T.}~\bibnamefont {Das}},\ }\href
  {https://doi.org/10.1103/RevModPhys.88.021004} {\bibfield  {journal}
  {\bibinfo  {journal} {Rev. Mod. Phys.}\ }\textbf {\bibinfo {volume} {88}},\
  \bibinfo {pages} {021004} (\bibinfo {year} {2016})}\BibitemShut {NoStop}%
\bibitem [{\citenamefont {Armitage}\ \emph {et~al.}(2018)\citenamefont
  {Armitage}, \citenamefont {Mele},\ and\ \citenamefont
  {Vishwanath}}]{armitage_weyl_2018}%
  \BibitemOpen
  \bibfield  {author} {\bibinfo {author} {\bibfnamefont {N.}~\bibnamefont
  {Armitage}}, \bibinfo {author} {\bibfnamefont {E.}~\bibnamefont {Mele}},\
  and\ \bibinfo {author} {\bibfnamefont {A.}~\bibnamefont {Vishwanath}},\
  }\href {https://doi.org/10.1103/RevModPhys.90.015001} {\bibfield  {journal}
  {\bibinfo  {journal} {Rev. Mod. Phys.}\ }\textbf {\bibinfo {volume} {90}},\
  \bibinfo {pages} {015001} (\bibinfo {year} {2018})}\BibitemShut {NoStop}%
\bibitem [{\citenamefont {Wan}\ \emph {et~al.}(2011)\citenamefont {Wan},
  \citenamefont {Turner}, \citenamefont {Vishwanath},\ and\ \citenamefont
  {Savrasov}}]{wan_topological_2011}%
  \BibitemOpen
  \bibfield  {author} {\bibinfo {author} {\bibfnamefont {X.}~\bibnamefont
  {Wan}}, \bibinfo {author} {\bibfnamefont {A.~M.}\ \bibnamefont {Turner}},
  \bibinfo {author} {\bibfnamefont {A.}~\bibnamefont {Vishwanath}},\ and\
  \bibinfo {author} {\bibfnamefont {S.~Y.}\ \bibnamefont {Savrasov}},\ }\href
  {https://doi.org/10.1103/PhysRevB.83.205101} {\bibfield  {journal} {\bibinfo
  {journal} {Phys. Rev. B}\ }\textbf {\bibinfo {volume} {83}},\ \bibinfo
  {pages} {205101} (\bibinfo {year} {2011})}\BibitemShut {NoStop}%
\bibitem [{\citenamefont {Nagaosa}\ \emph {et~al.}(2020)\citenamefont
  {Nagaosa}, \citenamefont {Morimoto},\ and\ \citenamefont
  {Tokura}}]{nagaosa_transport_2020}%
  \BibitemOpen
  \bibfield  {author} {\bibinfo {author} {\bibfnamefont {N.}~\bibnamefont
  {Nagaosa}}, \bibinfo {author} {\bibfnamefont {T.}~\bibnamefont {Morimoto}},\
  and\ \bibinfo {author} {\bibfnamefont {Y.}~\bibnamefont {Tokura}},\ }\href
  {https://doi.org/10.1038/s41578-020-0208-y} {\bibfield  {journal} {\bibinfo
  {journal} {Nat Rev Mater}\ }\textbf {\bibinfo {volume} {5}},\ \bibinfo
  {pages} {621} (\bibinfo {year} {2020})}\BibitemShut {NoStop}%
\bibitem [{\citenamefont {Ilan}\ \emph {et~al.}(2020)\citenamefont {Ilan},
  \citenamefont {Grushin},\ and\ \citenamefont
  {Pikulin}}]{ilan_pseudo-electromagnetic_2020}%
  \BibitemOpen
  \bibfield  {author} {\bibinfo {author} {\bibfnamefont {R.}~\bibnamefont
  {Ilan}}, \bibinfo {author} {\bibfnamefont {A.~G.}\ \bibnamefont {Grushin}},\
  and\ \bibinfo {author} {\bibfnamefont {D.~I.}\ \bibnamefont {Pikulin}},\
  }\href {https://doi.org/10.1038/s42254-019-0121-8} {\bibfield  {journal}
  {\bibinfo  {journal} {Nat Rev Phys}\ }\textbf {\bibinfo {volume} {2}},\
  \bibinfo {pages} {29} (\bibinfo {year} {2020})}\BibitemShut {NoStop}%
\bibitem [{\citenamefont {Destraz}\ \emph {et~al.}(2020)\citenamefont
  {Destraz}, \citenamefont {Das}, \citenamefont {Tsirkin}, \citenamefont {Xu},
  \citenamefont {Neupert}, \citenamefont {Chang}, \citenamefont {Schilling},
  \citenamefont {Grushin}, \citenamefont {Kohlbrecher}, \citenamefont {Keller},
  \citenamefont {Puphal}, \citenamefont {Pomjakushina},\ and\ \citenamefont
  {White}}]{destraz_magnetism_2020}%
  \BibitemOpen
  \bibfield  {author} {\bibinfo {author} {\bibfnamefont {D.}~\bibnamefont
  {Destraz}}, \bibinfo {author} {\bibfnamefont {L.}~\bibnamefont {Das}},
  \bibinfo {author} {\bibfnamefont {S.~S.}\ \bibnamefont {Tsirkin}}, \bibinfo
  {author} {\bibfnamefont {Y.}~\bibnamefont {Xu}}, \bibinfo {author}
  {\bibfnamefont {T.}~\bibnamefont {Neupert}}, \bibinfo {author} {\bibfnamefont
  {J.}~\bibnamefont {Chang}}, \bibinfo {author} {\bibfnamefont
  {A.}~\bibnamefont {Schilling}}, \bibinfo {author} {\bibfnamefont {A.~G.}\
  \bibnamefont {Grushin}}, \bibinfo {author} {\bibfnamefont {J.}~\bibnamefont
  {Kohlbrecher}}, \bibinfo {author} {\bibfnamefont {L.}~\bibnamefont {Keller}},
  \bibinfo {author} {\bibfnamefont {P.}~\bibnamefont {Puphal}}, \bibinfo
  {author} {\bibfnamefont {E.}~\bibnamefont {Pomjakushina}},\ and\ \bibinfo
  {author} {\bibfnamefont {J.~S.}\ \bibnamefont {White}},\ }\href
  {https://doi.org/10.1038/s41535-019-0207-7} {\bibfield  {journal} {\bibinfo
  {journal} {npj Quantum Mater.}\ }\textbf {\bibinfo {volume} {5}},\ \bibinfo
  {pages} {5} (\bibinfo {year} {2020})}\BibitemShut {NoStop}%
\bibitem [{\citenamefont {Yuan}\ \emph {et~al.}(2020)\citenamefont {Yuan},
  \citenamefont {Zhang}, \citenamefont {Zhang}, \citenamefont {Yan},
  \citenamefont {Lyu}, \citenamefont {Zhang}, \citenamefont {Li}, \citenamefont
  {Song}, \citenamefont {Zhao}, \citenamefont {Leng}, \citenamefont {Ozerov},
  \citenamefont {Chen}, \citenamefont {Wang}, \citenamefont {Shi},
  \citenamefont {Yan},\ and\ \citenamefont {Xiu}}]{yuan_discovery_2020}%
  \BibitemOpen
  \bibfield  {author} {\bibinfo {author} {\bibfnamefont {X.}~\bibnamefont
  {Yuan}}, \bibinfo {author} {\bibfnamefont {C.}~\bibnamefont {Zhang}},
  \bibinfo {author} {\bibfnamefont {Y.}~\bibnamefont {Zhang}}, \bibinfo
  {author} {\bibfnamefont {Z.}~\bibnamefont {Yan}}, \bibinfo {author}
  {\bibfnamefont {T.}~\bibnamefont {Lyu}}, \bibinfo {author} {\bibfnamefont
  {M.}~\bibnamefont {Zhang}}, \bibinfo {author} {\bibfnamefont
  {Z.}~\bibnamefont {Li}}, \bibinfo {author} {\bibfnamefont {C.}~\bibnamefont
  {Song}}, \bibinfo {author} {\bibfnamefont {M.}~\bibnamefont {Zhao}}, \bibinfo
  {author} {\bibfnamefont {P.}~\bibnamefont {Leng}}, \bibinfo {author}
  {\bibfnamefont {M.}~\bibnamefont {Ozerov}}, \bibinfo {author} {\bibfnamefont
  {X.}~\bibnamefont {Chen}}, \bibinfo {author} {\bibfnamefont {N.}~\bibnamefont
  {Wang}}, \bibinfo {author} {\bibfnamefont {Y.}~\bibnamefont {Shi}}, \bibinfo
  {author} {\bibfnamefont {H.}~\bibnamefont {Yan}},\ and\ \bibinfo {author}
  {\bibfnamefont {F.}~\bibnamefont {Xiu}},\ }\href
  {https://doi.org/10.1038/s41467-020-14749-4} {\bibfield  {journal} {\bibinfo
  {journal} {Nat Commun}\ }\textbf {\bibinfo {volume} {11}},\ \bibinfo {pages}
  {1259} (\bibinfo {year} {2020})}\BibitemShut {NoStop}%
\bibitem [{\citenamefont {Suzuki}\ \emph {et~al.}(2019)\citenamefont {Suzuki},
  \citenamefont {Savary}, \citenamefont {Liu}, \citenamefont {Lynn},
  \citenamefont {Balents},\ and\ \citenamefont
  {Checkelsky}}]{suzuki_singular_2019}%
  \BibitemOpen
  \bibfield  {author} {\bibinfo {author} {\bibfnamefont {T.}~\bibnamefont
  {Suzuki}}, \bibinfo {author} {\bibfnamefont {L.}~\bibnamefont {Savary}},
  \bibinfo {author} {\bibfnamefont {J.-P.}\ \bibnamefont {Liu}}, \bibinfo
  {author} {\bibfnamefont {J.~W.}\ \bibnamefont {Lynn}}, \bibinfo {author}
  {\bibfnamefont {L.}~\bibnamefont {Balents}},\ and\ \bibinfo {author}
  {\bibfnamefont {J.~G.}\ \bibnamefont {Checkelsky}},\ }\href
  {https://doi.org/10.1126/science.aat0348} {\bibfield  {journal} {\bibinfo
  {journal} {Science}\ }\textbf {\bibinfo {volume} {365}},\ \bibinfo {pages}
  {377} (\bibinfo {year} {2019})}\BibitemShut {NoStop}%
\bibitem [{\citenamefont {Yang}\ \emph {et~al.}(2021)\citenamefont {Yang},
  \citenamefont {Singh}, \citenamefont {Gaudet}, \citenamefont {Lu},
  \citenamefont {Huang}, \citenamefont {Chiu}, \citenamefont {Huang},
  \citenamefont {Wang}, \citenamefont {Bahrami}, \citenamefont {Xu},
  \citenamefont {Franklin}, \citenamefont {Sochnikov}, \citenamefont {Graf},
  \citenamefont {Xu}, \citenamefont {Zhao}, \citenamefont {Hoffman},
  \citenamefont {Lin}, \citenamefont {Torchinsky}, \citenamefont {Broholm},
  \citenamefont {Bansil},\ and\ \citenamefont
  {Tafti}}]{yang_noncollinear_2021}%
  \BibitemOpen
  \bibfield  {author} {\bibinfo {author} {\bibfnamefont {H.-Y.}\ \bibnamefont
  {Yang}}, \bibinfo {author} {\bibfnamefont {B.}~\bibnamefont {Singh}},
  \bibinfo {author} {\bibfnamefont {J.}~\bibnamefont {Gaudet}}, \bibinfo
  {author} {\bibfnamefont {B.}~\bibnamefont {Lu}}, \bibinfo {author}
  {\bibfnamefont {C.-Y.}\ \bibnamefont {Huang}}, \bibinfo {author}
  {\bibfnamefont {W.-C.}\ \bibnamefont {Chiu}}, \bibinfo {author}
  {\bibfnamefont {S.-M.}\ \bibnamefont {Huang}}, \bibinfo {author}
  {\bibfnamefont {B.}~\bibnamefont {Wang}}, \bibinfo {author} {\bibfnamefont
  {F.}~\bibnamefont {Bahrami}}, \bibinfo {author} {\bibfnamefont
  {B.}~\bibnamefont {Xu}}, \bibinfo {author} {\bibfnamefont {J.}~\bibnamefont
  {Franklin}}, \bibinfo {author} {\bibfnamefont {I.}~\bibnamefont {Sochnikov}},
  \bibinfo {author} {\bibfnamefont {D.~E.}\ \bibnamefont {Graf}}, \bibinfo
  {author} {\bibfnamefont {G.}~\bibnamefont {Xu}}, \bibinfo {author}
  {\bibfnamefont {Y.}~\bibnamefont {Zhao}}, \bibinfo {author} {\bibfnamefont
  {C.~M.}\ \bibnamefont {Hoffman}}, \bibinfo {author} {\bibfnamefont
  {H.}~\bibnamefont {Lin}}, \bibinfo {author} {\bibfnamefont {D.~H.}\
  \bibnamefont {Torchinsky}}, \bibinfo {author} {\bibfnamefont {C.~L.}\
  \bibnamefont {Broholm}}, \bibinfo {author} {\bibfnamefont {A.}~\bibnamefont
  {Bansil}},\ and\ \bibinfo {author} {\bibfnamefont {F.}~\bibnamefont
  {Tafti}},\ }\href {https://doi.org/10.1103/PhysRevB.103.115143} {\bibfield
  {journal} {\bibinfo  {journal} {Phys. Rev. B}\ }\textbf {\bibinfo {volume}
  {103}},\ \bibinfo {pages} {115143} (\bibinfo {year} {2021})}\BibitemShut
  {NoStop}%
\bibitem [{\citenamefont {Xu}\ \emph {et~al.}(2021)\citenamefont {Xu},
  \citenamefont {Franklin}, \citenamefont {Jayacody}, \citenamefont {Yang},
  \citenamefont {Tafti},\ and\ \citenamefont {Sochnikov}}]{xu_picoscale_2020}%
  \BibitemOpen
  \bibfield  {author} {\bibinfo {author} {\bibfnamefont {B.}~\bibnamefont
  {Xu}}, \bibinfo {author} {\bibfnamefont {J.}~\bibnamefont {Franklin}},
  \bibinfo {author} {\bibfnamefont {A.}~\bibnamefont {Jayacody}}, \bibinfo
  {author} {\bibfnamefont {H.-Y.}\ \bibnamefont {Yang}}, \bibinfo {author}
  {\bibfnamefont {F.}~\bibnamefont {Tafti}},\ and\ \bibinfo {author}
  {\bibfnamefont {I.}~\bibnamefont {Sochnikov}},\ }\bibfield  {journal}
  {\bibinfo  {journal} {Advanced Quantum Technologies}\ }\href
  {https://doi.org/https://doi.org/10.1002/qute.202000101}
  {https://doi.org/10.1002/qute.202000101} (\bibinfo {year} {2021})\BibitemShut
  {NoStop}%
\bibitem [{\citenamefont {Liu}\ \emph {et~al.}(2013)\citenamefont {Liu},
  \citenamefont {Ye},\ and\ \citenamefont {Qi}}]{liu_chiral_2013}%
  \BibitemOpen
  \bibfield  {author} {\bibinfo {author} {\bibfnamefont {C.-X.}\ \bibnamefont
  {Liu}}, \bibinfo {author} {\bibfnamefont {P.}~\bibnamefont {Ye}},\ and\
  \bibinfo {author} {\bibfnamefont {X.-L.}\ \bibnamefont {Qi}},\ }\href
  {https://doi.org/10.1103/PhysRevB.87.235306} {\bibfield  {journal} {\bibinfo
  {journal} {Phys. Rev. B}\ }\textbf {\bibinfo {volume} {87}},\ \bibinfo
  {pages} {235306} (\bibinfo {year} {2013})}\BibitemShut {NoStop}%
\bibitem [{\citenamefont {Shapourian}\ \emph {et~al.}(2015)\citenamefont
  {Shapourian}, \citenamefont {Hughes},\ and\ \citenamefont
  {Ryu}}]{shapourian_viscoelastic_2015}%
  \BibitemOpen
  \bibfield  {author} {\bibinfo {author} {\bibfnamefont {H.}~\bibnamefont
  {Shapourian}}, \bibinfo {author} {\bibfnamefont {T.~L.}\ \bibnamefont
  {Hughes}},\ and\ \bibinfo {author} {\bibfnamefont {S.}~\bibnamefont {Ryu}},\
  }\href {https://doi.org/10.1103/PhysRevB.92.165131} {\bibfield  {journal}
  {\bibinfo  {journal} {Phys. Rev. B}\ }\textbf {\bibinfo {volume} {92}},\
  \bibinfo {pages} {165131} (\bibinfo {year} {2015})}\BibitemShut {NoStop}%
\bibitem [{\citenamefont {Cortijo}\ \emph {et~al.}(2015)\citenamefont
  {Cortijo}, \citenamefont {Ferreirós}, \citenamefont {Landsteiner},\ and\
  \citenamefont {Vozmediano}}]{cortijo_elastic_2015}%
  \BibitemOpen
  \bibfield  {author} {\bibinfo {author} {\bibfnamefont {A.}~\bibnamefont
  {Cortijo}}, \bibinfo {author} {\bibfnamefont {Y.}~\bibnamefont {Ferreirós}},
  \bibinfo {author} {\bibfnamefont {K.}~\bibnamefont {Landsteiner}},\ and\
  \bibinfo {author} {\bibfnamefont {M.~A.}\ \bibnamefont {Vozmediano}},\ }\href
  {https://doi.org/10.1103/PhysRevLett.115.177202} {\bibfield  {journal}
  {\bibinfo  {journal} {Phys. Rev. Lett.}\ }\textbf {\bibinfo {volume} {115}},\
  \bibinfo {pages} {177202} (\bibinfo {year} {2015})}\BibitemShut {NoStop}%
\bibitem [{\citenamefont {Araki}(2020)}]{araki_magnetic_2020}%
  \BibitemOpen
  \bibfield  {author} {\bibinfo {author} {\bibfnamefont {Y.}~\bibnamefont
  {Araki}},\ }\href {https://doi.org/https://doi.org/10.1002/andp.201900287}
  {\bibfield  {journal} {\bibinfo  {journal} {Annalen der Physik}\ }\textbf
  {\bibinfo {volume} {532}},\ \bibinfo {pages} {1900287} (\bibinfo {year}
  {2020})}\BibitemShut {NoStop}%
\bibitem [{\citenamefont {Hannukainen}\ \emph {et~al.}(2020)\citenamefont
  {Hannukainen}, \citenamefont {Ferreiros}, \citenamefont {Cortijo},\ and\
  \citenamefont {Bardarson}}]{hannukainen_axial_2020}%
  \BibitemOpen
  \bibfield  {author} {\bibinfo {author} {\bibfnamefont {J.~D.}\ \bibnamefont
  {Hannukainen}}, \bibinfo {author} {\bibfnamefont {Y.}~\bibnamefont
  {Ferreiros}}, \bibinfo {author} {\bibfnamefont {A.}~\bibnamefont {Cortijo}},\
  and\ \bibinfo {author} {\bibfnamefont {J.~H.}\ \bibnamefont {Bardarson}},\
  }\href {https://doi.org/10.1103/PhysRevB.102.241401} {\bibfield  {journal}
  {\bibinfo  {journal} {Phys. Rev. B}\ }\textbf {\bibinfo {volume} {102}},\
  \bibinfo {pages} {241401} (\bibinfo {year} {2020})}\BibitemShut {NoStop}%
\bibitem [{\citenamefont {Zyuzin}\ and\ \citenamefont
  {Zyuzin}(2015)}]{zyuzin_chiral_2015}%
  \BibitemOpen
  \bibfield  {author} {\bibinfo {author} {\bibfnamefont {A.~A.}\ \bibnamefont
  {Zyuzin}}\ and\ \bibinfo {author} {\bibfnamefont {V.~A.}\ \bibnamefont
  {Zyuzin}},\ }\href@noop {} {\bibfield  {journal} {\bibinfo  {journal} {Phys.
  Rev. B}\ ,\ \bibinfo {pages} {4}} (\bibinfo {year} {2015})}\BibitemShut
  {NoStop}%
\bibitem [{\citenamefont {Lux}\ \emph {et~al.}(2018)\citenamefont {Lux},
  \citenamefont {Freimuth}, \citenamefont {Blügel},\ and\ \citenamefont
  {Mokrousov}}]{lux_engineering_2018}%
  \BibitemOpen
  \bibfield  {author} {\bibinfo {author} {\bibfnamefont {F.~R.}\ \bibnamefont
  {Lux}}, \bibinfo {author} {\bibfnamefont {F.}~\bibnamefont {Freimuth}},
  \bibinfo {author} {\bibfnamefont {S.}~\bibnamefont {Blügel}},\ and\ \bibinfo
  {author} {\bibfnamefont {Y.}~\bibnamefont {Mokrousov}},\ }\href
  {https://doi.org/10.1038/s42005-018-0055-y} {\bibfield  {journal} {\bibinfo
  {journal} {Commun Phys}\ }\textbf {\bibinfo {volume} {1}},\ \bibinfo {pages}
  {60} (\bibinfo {year} {2018})}\BibitemShut {NoStop}%
\bibitem [{\citenamefont {Tchoumakov}\ \emph {et~al.}(2017)\citenamefont
  {Tchoumakov}, \citenamefont {Civelli},\ and\ \citenamefont
  {Goerbig}}]{tchoumakov_magnetic_2017}%
  \BibitemOpen
  \bibfield  {author} {\bibinfo {author} {\bibfnamefont {S.}~\bibnamefont
  {Tchoumakov}}, \bibinfo {author} {\bibfnamefont {M.}~\bibnamefont
  {Civelli}},\ and\ \bibinfo {author} {\bibfnamefont {M.~O.}\ \bibnamefont
  {Goerbig}},\ }\href {https://doi.org/10.1103/PhysRevB.95.125306} {\bibfield
  {journal} {\bibinfo  {journal} {Phys. Rev. B}\ }\textbf {\bibinfo {volume}
  {95}},\ \bibinfo {pages} {125306} (\bibinfo {year} {2017})}\BibitemShut
  {NoStop}%
\bibitem [{\citenamefont {Liang}\ and\ \citenamefont
  {Ojanen}(2020)}]{liang_topological_2020}%
  \BibitemOpen
  \bibfield  {author} {\bibinfo {author} {\bibfnamefont {L.}~\bibnamefont
  {Liang}}\ and\ \bibinfo {author} {\bibfnamefont {T.}~\bibnamefont {Ojanen}},\
  }\href {https://doi.org/10.1103/PhysRevResearch.2.022016} {\bibfield
  {journal} {\bibinfo  {journal} {Phys. Rev. Research}\ }\textbf {\bibinfo
  {volume} {2}},\ \bibinfo {pages} {022016} (\bibinfo {year}
  {2020})}\BibitemShut {NoStop}%
\bibitem [{\citenamefont {McCord}(2015)}]{mccord_progress_2015}%
  \BibitemOpen
  \bibfield  {author} {\bibinfo {author} {\bibfnamefont {J.}~\bibnamefont
  {McCord}},\ }\href {https://doi.org/10.1088/0022-3727/48/33/333001}
  {\bibfield  {journal} {\bibinfo  {journal} {J. Phys. D: Appl. Phys.}\
  }\textbf {\bibinfo {volume} {48}},\ \bibinfo {pages} {333001} (\bibinfo
  {year} {2015})}\BibitemShut {NoStop}%
\bibitem [{\citenamefont {Qiu}\ and\ \citenamefont
  {Bader}(2000)}]{qiu_surface_2000}%
  \BibitemOpen
  \bibfield  {author} {\bibinfo {author} {\bibfnamefont {Z.~Q.}\ \bibnamefont
  {Qiu}}\ and\ \bibinfo {author} {\bibfnamefont {S.~D.}\ \bibnamefont
  {Bader}},\ }\href {https://doi.org/10.1063/1.1150496} {\bibfield  {journal}
  {\bibinfo  {journal} {Review of Scientific Instruments}\ }\textbf {\bibinfo
  {volume} {71}},\ \bibinfo {pages} {1243} (\bibinfo {year}
  {2000})}\BibitemShut {NoStop}%
\bibitem [{\citenamefont {Stupakiewicz}\ \emph {et~al.}(2014)\citenamefont
  {Stupakiewicz}, \citenamefont {Chizhik}, \citenamefont {Tekielak},
  \citenamefont {Zhukov}, \citenamefont {Gonzalez},\ and\ \citenamefont
  {Maziewski}}]{stupakiewicz_direct_2014}%
  \BibitemOpen
  \bibfield  {author} {\bibinfo {author} {\bibfnamefont {A.}~\bibnamefont
  {Stupakiewicz}}, \bibinfo {author} {\bibfnamefont {A.}~\bibnamefont
  {Chizhik}}, \bibinfo {author} {\bibfnamefont {M.}~\bibnamefont {Tekielak}},
  \bibinfo {author} {\bibfnamefont {A.}~\bibnamefont {Zhukov}}, \bibinfo
  {author} {\bibfnamefont {J.}~\bibnamefont {Gonzalez}},\ and\ \bibinfo
  {author} {\bibfnamefont {A.}~\bibnamefont {Maziewski}},\ }\href
  {https://doi.org/10.1063/1.4896758} {\bibfield  {journal} {\bibinfo
  {journal} {Review of Scientific Instruments}\ }\textbf {\bibinfo {volume}
  {85}},\ \bibinfo {pages} {103702} (\bibinfo {year} {2014})}\BibitemShut
  {NoStop}%
\bibitem [{\citenamefont {Rave}\ \emph {et~al.}(1987)\citenamefont {Rave},
  \citenamefont {Schäfer},\ and\ \citenamefont
  {Hubert}}]{rave_quantitative_1987}%
  \BibitemOpen
  \bibfield  {author} {\bibinfo {author} {\bibfnamefont {W.}~\bibnamefont
  {Rave}}, \bibinfo {author} {\bibfnamefont {R.}~\bibnamefont {Schäfer}},\
  and\ \bibinfo {author} {\bibfnamefont {A.}~\bibnamefont {Hubert}},\ }\href
  {https://doi.org/10.1016/0304-8853(87)90304-0} {\bibfield  {journal}
  {\bibinfo  {journal} {Journal of Magnetism and Magnetic Materials}\ }\textbf
  {\bibinfo {volume} {65}},\ \bibinfo {pages} {7} (\bibinfo {year}
  {1987})}\BibitemShut {NoStop}%
\bibitem [{\citenamefont {Yang}\ and\ \citenamefont
  {Scheinfein}(1993)}]{yang_combined_1993}%
  \BibitemOpen
  \bibfield  {author} {\bibinfo {author} {\bibfnamefont {Z.~J.}\ \bibnamefont
  {Yang}}\ and\ \bibinfo {author} {\bibfnamefont {M.~R.}\ \bibnamefont
  {Scheinfein}},\ }\href {https://doi.org/10.1063/1.355081} {\bibfield
  {journal} {\bibinfo  {journal} {Journal of Applied Physics}\ }\textbf
  {\bibinfo {volume} {74}},\ \bibinfo {pages} {6810} (\bibinfo {year}
  {1993})}\BibitemShut {NoStop}%
\bibitem [{\citenamefont {Daboo}\ \emph {et~al.}(1993)\citenamefont {Daboo},
  \citenamefont {Bland}, \citenamefont {Hicken}, \citenamefont {Ives},
  \citenamefont {Baird},\ and\ \citenamefont {Walker}}]{daboo_vectorial_1993}%
  \BibitemOpen
  \bibfield  {author} {\bibinfo {author} {\bibfnamefont {C.}~\bibnamefont
  {Daboo}}, \bibinfo {author} {\bibfnamefont {J.~A.~C.}\ \bibnamefont {Bland}},
  \bibinfo {author} {\bibfnamefont {R.~J.}\ \bibnamefont {Hicken}}, \bibinfo
  {author} {\bibfnamefont {A.~J.~R.}\ \bibnamefont {Ives}}, \bibinfo {author}
  {\bibfnamefont {M.~J.}\ \bibnamefont {Baird}},\ and\ \bibinfo {author}
  {\bibfnamefont {M.~J.}\ \bibnamefont {Walker}},\ }\href
  {https://doi.org/10.1103/PhysRevB.47.11852} {\bibfield  {journal} {\bibinfo
  {journal} {Phys. Rev. B}\ }\textbf {\bibinfo {volume} {47}},\ \bibinfo
  {pages} {11852} (\bibinfo {year} {1993})}\BibitemShut {NoStop}%
\bibitem [{\citenamefont {Ding}\ \emph {et~al.}(2000)\citenamefont {Ding},
  \citenamefont {Pütter}, \citenamefont {Oepen},\ and\ \citenamefont
  {Kirschner}}]{ding_experimental_2000}%
  \BibitemOpen
  \bibfield  {author} {\bibinfo {author} {\bibfnamefont {H.}~\bibnamefont
  {Ding}}, \bibinfo {author} {\bibfnamefont {S.}~\bibnamefont {Pütter}},
  \bibinfo {author} {\bibfnamefont {H.}~\bibnamefont {Oepen}},\ and\ \bibinfo
  {author} {\bibfnamefont {J.}~\bibnamefont {Kirschner}},\ }\href
  {https://doi.org/10.1016/S0304-8853(99)00790-8} {\bibfield  {journal}
  {\bibinfo  {journal} {Journal of Magnetism and Magnetic Materials}\ }\textbf
  {\bibinfo {volume} {212}},\ \bibinfo {pages} {5} (\bibinfo {year}
  {2000})}\BibitemShut {NoStop}%
\bibitem [{\citenamefont {Ullah}\ \emph {et~al.}(2019)\citenamefont {Ullah},
  \citenamefont {Balamurugan}, \citenamefont {Zhang}, \citenamefont
  {Valloppilly}, \citenamefont {Li}, \citenamefont {Pahari}, \citenamefont
  {Yue}, \citenamefont {Sokolov}, \citenamefont {Sellmyer},\ and\ \citenamefont
  {Skomski}}]{ullah_crystal_2019}%
  \BibitemOpen
  \bibfield  {author} {\bibinfo {author} {\bibfnamefont {A.}~\bibnamefont
  {Ullah}}, \bibinfo {author} {\bibfnamefont {B.}~\bibnamefont {Balamurugan}},
  \bibinfo {author} {\bibfnamefont {W.}~\bibnamefont {Zhang}}, \bibinfo
  {author} {\bibfnamefont {S.}~\bibnamefont {Valloppilly}}, \bibinfo {author}
  {\bibfnamefont {X.-Z.}\ \bibnamefont {Li}}, \bibinfo {author} {\bibfnamefont
  {R.}~\bibnamefont {Pahari}}, \bibinfo {author} {\bibfnamefont {L.-P.}\
  \bibnamefont {Yue}}, \bibinfo {author} {\bibfnamefont {A.}~\bibnamefont
  {Sokolov}}, \bibinfo {author} {\bibfnamefont {D.~J.}\ \bibnamefont
  {Sellmyer}},\ and\ \bibinfo {author} {\bibfnamefont {R.}~\bibnamefont
  {Skomski}},\ }\href {https://doi.org/10.1109/TMAG.2018.2890028} {\bibfield
  {journal} {\bibinfo  {journal} {IEEE Trans. Magn.}\ }\textbf {\bibinfo
  {volume} {55}},\ \bibinfo {pages} {1} (\bibinfo {year} {2019})}\BibitemShut
  {NoStop}%
\end{thebibliography}%
\end{document}


\maketitle

\section{Experimental Set-up and Balanced Optical Bridge Detector}

\begin{figure}[H]
\centering
\includegraphics[width=\linewidth]{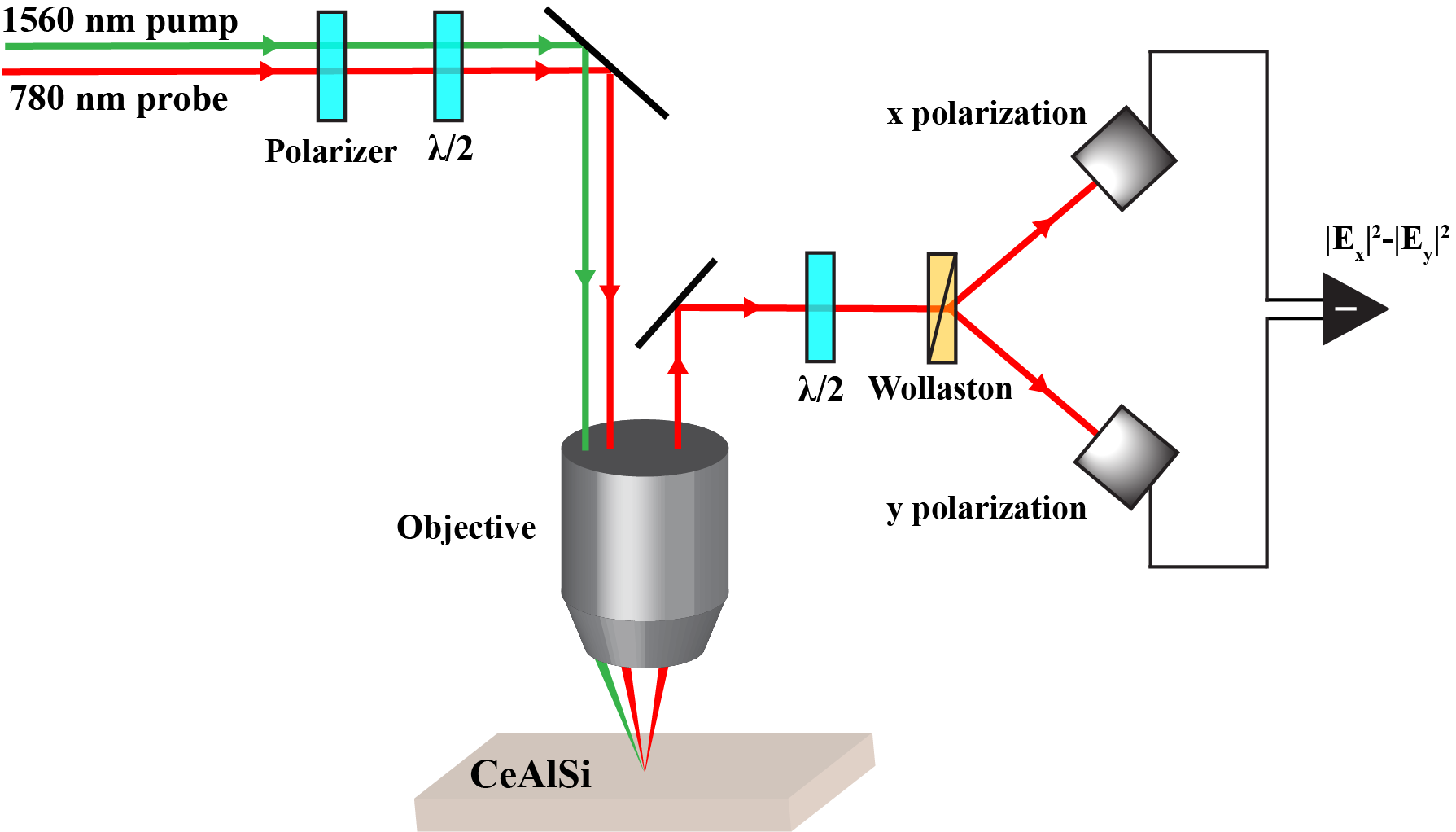}
\caption{Schematic of vector MOKE microscopy and balanced optical bridge detector.\label{fig:sup}}
\end{figure}

A schematic of the optical set-up is shown in Figure S1. The polarization of the 780 nm probe beam is set by a combination of a polarizer and a rotatable half-wave plate. The 1560 nm pump beam is modulated at 2.5 kHz with a chopper, to provide thermal modulation. Both beams are focused onto the sample surface with a 10x objective. The measured vector MOKE signal  $\Theta$ is independent from the pump beam polarization. The reflected 1560 nm pump beam is blocked by a color filter, and the silicon-based photodiodes are not sensitive to the 1560 nm light. The Wollaston prism spatially separates the reflected probe beam to horizontal (x) and vertical (y) polarization components, which are detected by a pair of unbiased photodiodes. Then the difference of the signals from two photodiodes is read as the signal $\Theta$. When the x and y polarization components are equal, the signal  $\Theta$ will be zero, so that the two photodiodes are balanced. The balancing is achieved by tuning another half-wave plate before the Wollaston prism above the transition temperature $T_{c}$.

\section{Macroscopic Vector MOKE Model}

To gain more insights into the polarization dependence of different MOKE components, we construct a macroscopic MOKE model based on an argument of the dielectric tensor. We attribute the response of a magnetic medium into the asymmetric parts of the dielectric tensor, since all the symmetric parts can be diagonalized and contribute nothing to the rotation of polarization. The argument makes the dielectric tensor  to have the form \cite{qiu_surface_2000},

\begin{equation}
\tilde{\epsilon} = \epsilon \begin{bmatrix} 1 & iQM_{z} & -iQM_{y} \\ -iQM_{z} & 1 & iQM_{x} \\ iQM_{y} & -iQM_{x} & 1 \end{bmatrix},
\end{equation}
where $Q$ is the magneto-optical constant, $M_{z}$ is the out-of-plane magnetization, $M_{y}$ is the in-plane magnetization lying in the incident plane and $M_{x}$ is the in-plane magnetization normal to the incident plane. By substituting the dielectric tensor into the Maxwell equations and applying boundary conditions on the air-magnet interface \cite{qiu_surface_2000}, we calculate the reflection tensor $R$ at the air-magnet interface.

\begin{equation}
R=\begin{bmatrix} r_{ss} & r_{sp} \\ r_{ps} & r_{pp} \\ \end{bmatrix}
=\begin{bmatrix} r & \beta M_{y}-\gamma M_{z} \\ -\beta M_{y}-\gamma M_{z} & -r-2\alpha M_{x} \\ \end{bmatrix},
\end{equation}
where
\begin{align}
&r=\frac{n_{1}\cos\theta_{1}-n_{2}\cos\theta_{2}}{n_{1}\cos\theta_{1}+n_{2}\cos\theta_{2}},
\\&\alpha=\frac{Qn_{1}n_{2}\cos\theta_{1}\sin\theta_{2}}{(n_{1}\cos\theta_{2}+n_{2}\cos\theta_{1})^{2}},
\\& \beta=\frac{Qn_{1}n_{2}\cos\theta_{1}\sin\theta_{2}}{(n_{2}\cos\theta_{1}+n_{1}\cos\theta_{2})(n_{1}\cos\theta_{1}+n_{2}\cos\theta_{2})\cos\theta_{2}},
\\& \gamma=\frac{Qn_{1}n_{2}\cos\theta_{1}\cos\theta_{2}}{(n_{2}\cos\theta_{1}+n_{1}\cos\theta_{2})(n_{1}\cos\theta_{1}+n_{2}\cos\theta_{2})\cos\theta_{2}}.
\end{align}
$n_{1}$ is the refractive index of air, $\theta_{1}$ is the incident angle, $n_{2}$ is the refractive index of the magnetic material and $\theta_{2}$ satisfies $n_{2}\sin\ \theta_{2}=n_{1}\sin\ \theta_{1}$. Both $M_{z}$ and $M_{y}$ show up in the off-diagonal terms, inducing Kerr rotations. The $M_{y}$ terms have opposite signs in $r_{sp}$ and $r_{ps}$, indicating the polarization dependence. The $M_{x}$ term is the most subtle one, which is often ignored in previous papers\cite{qiu_surface_2000,stupakiewicz_direct_2014,ding_experimental_2000,daboo_vectorial_1993}. $M_{x}$ generates optical birefringence, which cannot be detected by either $s$- or $p$-polarized beam, but can be detected when the incident polarization is rotated by $\pm$45° with respect to the plane of incidence.

Although the incident beam is oblique, the incident angle $\theta_{1}$ is still tiny, so $\theta_{1}<<1$, $\theta_{2}<<1$. Expanding the reflection tensor to the first order of $\theta_{1}$ and $\theta_{2}$, we have
\begin{align}
&r=\frac{n_{1}-n_{2}}{n_{1}+n_{2}},
\\&\alpha=\beta\approx\frac{Qn_{1}^{2}\theta_{1}}{(n_{1}+n_{2})^{2}},
\\& \gamma\approx\frac{Qn_{1}n_{2}}{(n_{1}+n_{2})^{2}}.
\end{align}
The prefactors of $M_{x}$ and $M_{y}$ are approximately equal to each other, which paves the way to measure maps of in-plane magnetization. 

We can use the reflection tensor to calculate the MOKE signal measured with the set-up shown in Figure S1. The incident beam can be represented by a vector $E_{i}=\begin{bmatrix} \cos\phi \\ \sin\phi \end{bmatrix}$, where $\phi$ is the angle of incident polarization. $\phi=0$ corresponds to the $s$-polarized beam and $\phi=\pi/2$ is the $p$-polarized beam. The combination of the second half-wave plate and the Wollaston prism can be expressed as a Jones matrix $W$.
\begin{equation}
W=\frac{1}{\sqrt{2}}\begin{bmatrix} \cos\phi-\sin\phi & -\cos\phi-\sin\phi \\ \cos\phi+\sin\phi & \cos\phi-\sin\phi \\\end{bmatrix}
\end{equation}
The matrix $W$ can map the vector $\begin{bmatrix} r\cos\phi \\ -r\sin\phi \end{bmatrix}$ to $\frac{r}{\sqrt{2}}\begin{bmatrix} 1 \\ 1 \end{bmatrix}$, which corresponds to the balancing operation above $T_{c}$. Then the light field $E_{f}$ after the Wollaston prism can be calculated as:
\begin{equation}
\begin{aligned}
E_{f}&=WRE_{i}
\\&=\frac{1}{\sqrt{2}}\begin{bmatrix} \cos\phi-\sin\phi & -\cos\phi-\sin\phi \\ \cos\phi+\sin\phi & \cos\phi-\sin\phi \\\end{bmatrix}\begin{bmatrix} r & \beta M_{y}-\gamma M_{z} \\ -\beta M_{y}-\gamma M_{z} & -r-2\alpha M_{x} \\ \end{bmatrix}\begin{bmatrix} \cos\phi \\ \sin\phi \end{bmatrix}
\\&=\frac{1}{\sqrt{2}}\begin{bmatrix} r+\alpha M_{x}+\gamma M_{z}+(-\alpha M_{x}+\beta M_{y})\cos(2\phi)+(\alpha M_{x}+\beta M_{y})\sin(2\phi) 
\\ r+\alpha M_{x}-\gamma M_{z}-(\alpha M_{x}+\beta M_{y})\cos(2\phi)+(-\alpha M_{x}+\beta M_{y})\sin(2\phi)
\end{bmatrix}
\\&=\begin{bmatrix} E_{x} \\ E_{y} \end{bmatrix}.
\end{aligned}
\end{equation}
The signal $\Delta I$ measured by the balanced optical bridge detector is $\Delta I=|E_{x}|^{2}-|E_{y}|^{2}$. To the first order of $M_{x}$, $M_{y}$ and $M_{z}$,
\begin{equation}
\Delta I=2Re(r\gamma^{*}) M_{z}+2Re(r\beta^{*}) M_{y}\cos(2\phi)+2Re(r\alpha^{*}) M_{x}\sin(2\phi)
\end{equation}
For $s$-polarized incident beam, $\phi=0$,
\begin{equation}
\Delta I (\phi=0)=2Re(r\gamma^{*}) M_{z}+2Re(r\beta^{*}) M_{y}
\end{equation}
For $p$-polarized incident beam, $\phi=\pi/2$,
\begin{equation}
\Delta I (\phi=\pi/2)=2Re(r\gamma^{*}) M_{z}-2Re(r\beta^{*}) M_{y}
\end{equation}
For $\phi=\pi/4$, which can be called as $s$+$p$-polarized incident beam, 
\begin{equation}
\Delta I (\phi=\pi/4)=2Re(r\gamma^{*}) M_{z}+2Re(r\alpha^{*}) M_{x}
\end{equation}
Finally, the vector MOKE signal $\Theta$ can be expressed as
\begin{equation}
\begin{aligned}
&\Theta_{y}=\frac{1}{2}\left[\Delta I (\phi=0)-\Delta I (\phi=\pi/2)\right]=2Re(r\beta^{*})  M_{y}
\\&\Theta_{z}=\frac{1}{2}\left[\Delta I (\phi=0)+\Delta I (\phi=\pi/2)\right]=2Re(r\gamma^{*})  M_{z}
\\&\Theta_{x}=\Delta I (\phi=\pi/4)-\Theta_{z}=2Re(r\alpha^{*})  M_{x}
\end{aligned}
\end{equation}
This result is also supported by some previous papers\cite{rave_quantitative_1987,yang_combined_1993}.

\section{Thermal-modulated Vector MOKE}
In order to eliminate long term drifts and achieve microradian sensitivity, we overlap a 780 nm probe beam with a 1560 nm pump beam chopped at 2.5 kHz, as shown in Figure S1. Lock-in detection captures the thermal-modulated vector MOKE signal $\delta \boldsymbol{\Theta}$, so it is crucial to figure out the relation between $\delta \boldsymbol{\Theta}$ and the magnetization $\boldsymbol{M}$.

In CeAlSi, the in-plane components dominate the magnetization, so we focus on the analysis of the in-plane components $\boldsymbol{\Theta}_{\parallel}$ and $\boldsymbol{M}_{\parallel}$. Considering $\alpha \approx \beta$, $\boldsymbol{\Theta}_{\parallel}=2Re(r\alpha^{*})\boldsymbol{M_{\parallel}}$. Then we have
\begin{equation}
\delta \boldsymbol{\Theta}_{\parallel}=2\ \delta \left[Re(r\alpha^{*})\right]\cdot \boldsymbol{M}_{\parallel}+2Re(r\alpha^{*})\cdot\delta \boldsymbol{M}_{\parallel}.
\end{equation}
The first term is always along the direction of $\boldsymbol{M}_{\parallel}$, while the second term can be separated into two orthogonal components. The changes of both amplitude and direction of $\boldsymbol{M}_{\parallel}$contribute to $\delta \boldsymbol{M}_{\parallel}$. The change of amplitude $\delta |\boldsymbol{M}_{\parallel}|$ is along the direction of $\boldsymbol{M}_{\parallel}$, so we denote the longitudinal change as $\left(\delta\boldsymbol{M}_{\parallel}\right)_{L}$. The change induced by the rotation of $\boldsymbol{M}_{\parallel}$ is perpendicular to the direction of $\boldsymbol{M}_{\parallel}$, so we denote the transverse change as $\left(\delta \boldsymbol{M}_{\parallel}\right)_{T}$. Then we can write $\delta \Theta_{\parallel}$ as
\begin{equation}
\begin{aligned}
\delta \boldsymbol{\Theta}_{\parallel}
&=2\ \delta \left[Re(r\alpha^{*})\right]\cdot \boldsymbol{M}_{\parallel}+2Re(r\alpha^{*})\cdot\left(\delta \boldsymbol{M}_{\parallel}\right)_{L}+2Re(r\alpha^{*})\cdot\left(\delta \boldsymbol{M}_{\parallel}\right)_{T}
\\&=\left(\delta \boldsymbol{\Theta}_{\parallel}\right)_{L}+\left(\delta \boldsymbol{\Theta}_{\parallel}\right)_{T},
\end{aligned}
\end{equation}
where $\left(\delta \boldsymbol{\Theta}_{\parallel}\right)_{L}=2\ \delta \left[Re(r\alpha^{*})\right]\cdot \boldsymbol{M}_{\parallel}+2Re(r\alpha^{*})\cdot\left(\delta \boldsymbol{M}_{\parallel}\right)_{L}$ and $\left(\delta \boldsymbol{\Theta}_{\parallel}\right)_{T}=2Re(r\alpha^{*})\cdot\left(\delta \boldsymbol{M}_{\parallel}\right)_{T}$. $\left(\delta \boldsymbol{\Theta}_{\parallel}\right)_{L}$ is along the direction of $\boldsymbol{M}_{\parallel}$, while $\left(\delta \boldsymbol{\Theta}_{\parallel}\right)_{T}$ is perpendicular to the direction of $\boldsymbol{M}_{\parallel}$. The transverse component only comes from the rotation of $\boldsymbol{M}_{\parallel}$, so we have
\begin{equation}
\left(\delta \boldsymbol{\Theta}_{\parallel}\right)_{T}=2Re(r\alpha^{*})|\boldsymbol{M}_{\parallel}|\delta \epsilon=|\boldsymbol{\Theta}_{\parallel}|\delta\epsilon,
\end{equation}
where $\delta \epsilon$ is the variation of $\boldsymbol{M}_{\parallel}$ direction. We will estimate the value of $\left(\delta \boldsymbol{\Theta}_{\parallel}\right)_{T}$ below.

In Figure 2(b) of the main text, the direction of $\delta\boldsymbol{\Theta}_{\parallel}$ is plotted as a function of temperature, which is sensitive to the direction of $\boldsymbol{M}_{\parallel}$. Between 2 K and 6 K, the change of $\delta \boldsymbol{\Theta}_{\parallel}$ direction is very small. For example, from 2 K to 4 K, the direction of $\delta \boldsymbol{\Theta}_{\parallel}$ only changes by 1°. The temperature modulation $\delta T$ at 2 K is smaller than 2 K, because the domain patterns change from 2 K to 4 K in Figure 2(a) of the main text. Then it is reasonable to suppose $\delta \epsilon$ at 2 K is smaller than 1°. The value of $|\boldsymbol{\Theta}_{\parallel}|$ can be estimated via the vector MOKE measurement without thermal modulation. The vector MOKE signal $|\boldsymbol{\Theta}_{\parallel}|$ is smaller than the noise level without thermal modulation, which is $\sim$1 mrad. In this way, we have
\begin{equation}
\left(\delta \boldsymbol{\Theta}_{\parallel}\right)_{T}\leqslant 1 \ mrad*\frac{1}{180}\pi\approx 15\ \mu rad
\end{equation}
The value of $\delta\boldsymbol{\Theta}_{\parallel}$ is as large as 150 $\mu$rad at 2 K, which is 10 times larger than the upper bound of $\left(\delta \boldsymbol{\Theta}_{\parallel}\right)_{T}$, so the thermal-modulated MOKE signal $\delta \boldsymbol{\Theta}_{\parallel}$ is almost along the direction of $\boldsymbol{M}_{\parallel}$.

\section{Local Mirror Symmetry Breaking}
The chirality of the vertical domain walls requires the presence of a local Lifshitz invariant of the form, $DM_{z}\left(\nabla\times \boldsymbol{M}\right)_{z}$. The term requires the breaking of mirror symmetry. It is the noncollinear magnetic order of CeAlSi that allows the mirror symmetry breaking locally at the vertical domain walls.

Take the vertical domain wall between the yellow and purple domains at 2 K as an example. The magnetization in the yellow domain points to approximately 70° and the purple domain points to approximately 290°, which still obeys the mirror symmetry by viewing the net magnetization directions. The noncollinear order requires the magnetizations of the alternating layers span an angle $2\phi_{0}$. Without loss of generality, we can use $\phi_{0}$ = 25° extracted from Figure 2(c) in the main text. Then the alternating layers of the yellow domain will point to 45° and 95°, with 265° and 315° for the purple domain. There are two ways to arrange the four angles into alternating layers. As shown in Figure S2, Case I has 45° and 315° on the layer 1, with 95° and 265° on the layer 2. Case II has 95° and 315° on the layer 1, with 45° and 265° on the layer 2.
\begin{figure}[H]
\centering
\includegraphics[width=\linewidth]{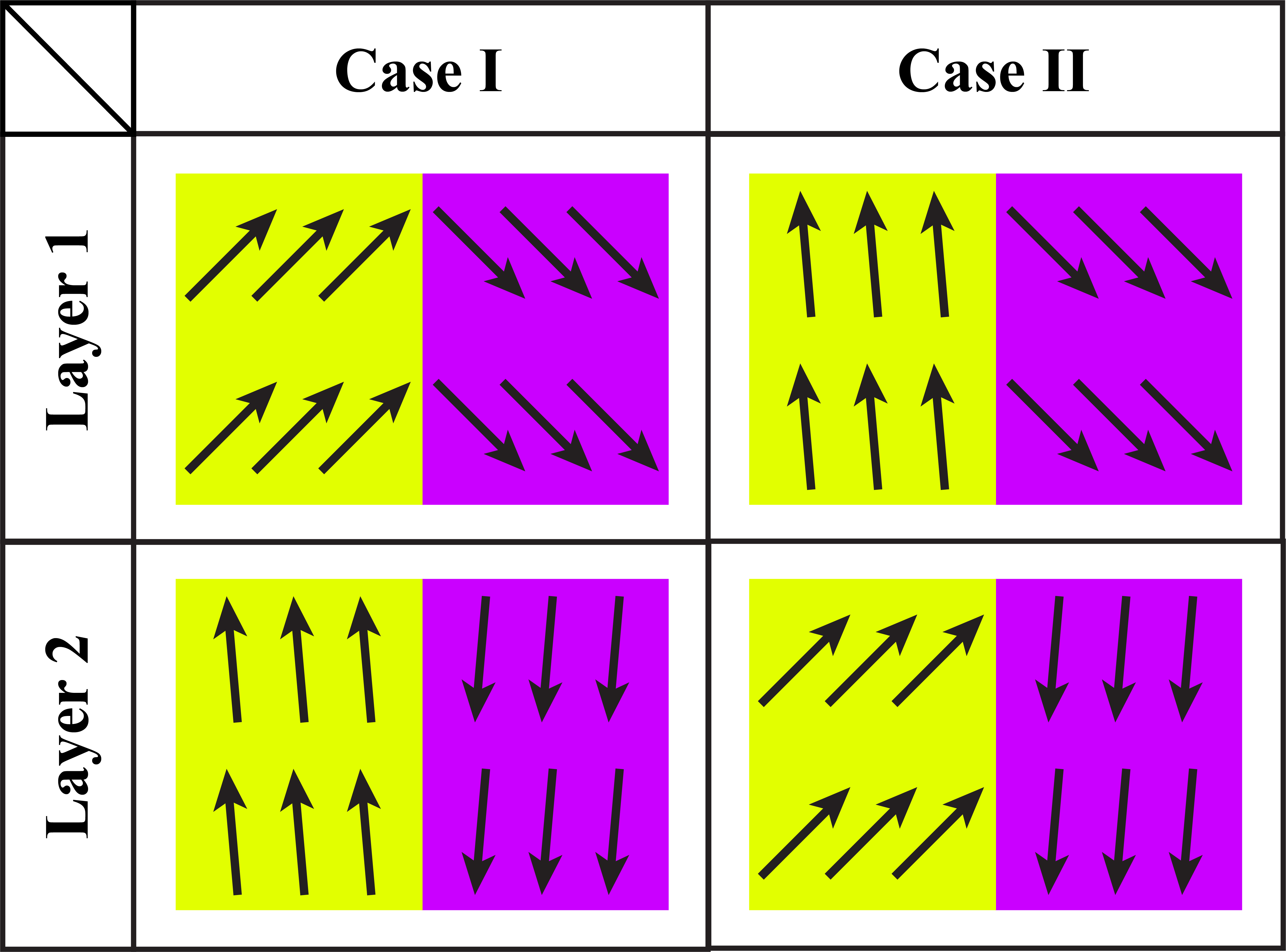}
\caption{Two ways to arrange magnetization directions in alternating layers.\label{fig:sup}}
\end{figure}
Both cases preserve mirror symmetry inside the domains, and have the same net magnetization map. Case II breaks the mirror symmetry locally at the domain wall in each layer, while the mirror symmetry is still preserved at the domain wall in Case I. If we denote the magnetization in the layer 1 as $\boldsymbol{M_{1}}$ and the layer 2 as $\boldsymbol{M_{2}}$, and calculate $\left(\boldsymbol{M_{1}}\times\boldsymbol{M_{2}}\right)_{z}$, we can find the subtle difference between Case I and II. In Case I, we have $\left(\boldsymbol{M_{1}}\times\boldsymbol{M_{2}}\right)_{z}>0$ for the yellow domain but $\left(\boldsymbol{M_{1}}\times\boldsymbol{M_{2}}\right)_{z}<0$ for the purple domain. As a contrast, Case II has $\left(\boldsymbol{M_{1}}\times\boldsymbol{M_{2}}\right)_{z}<0$ in both domains. The local mirror symmetry breaking requires $\left(\boldsymbol{M_{1}}\times\boldsymbol{M_{2}}\right)_{z}$ to have the same sign on both sides of the domain wall, which is allowed by minimizing the invariant $|\partial_{x}\left(\boldsymbol{M_{1}}\times\boldsymbol{M_{2}}\right)_{z}|^{2}$.

\newpage
\section{Vector MOKE Map Above the Transition Temperature}
The vector MOKE signal $\boldsymbol{\Theta}$ drops to approximately zero above the transition temperature. In Figure S3, the vector MOKE map captured at 9 K is shown. The noise dominates the map and no domains can be observed, which corresponds to the SQUID measurements.

\begin{figure}[H]
\centering
\includegraphics[width=\linewidth]{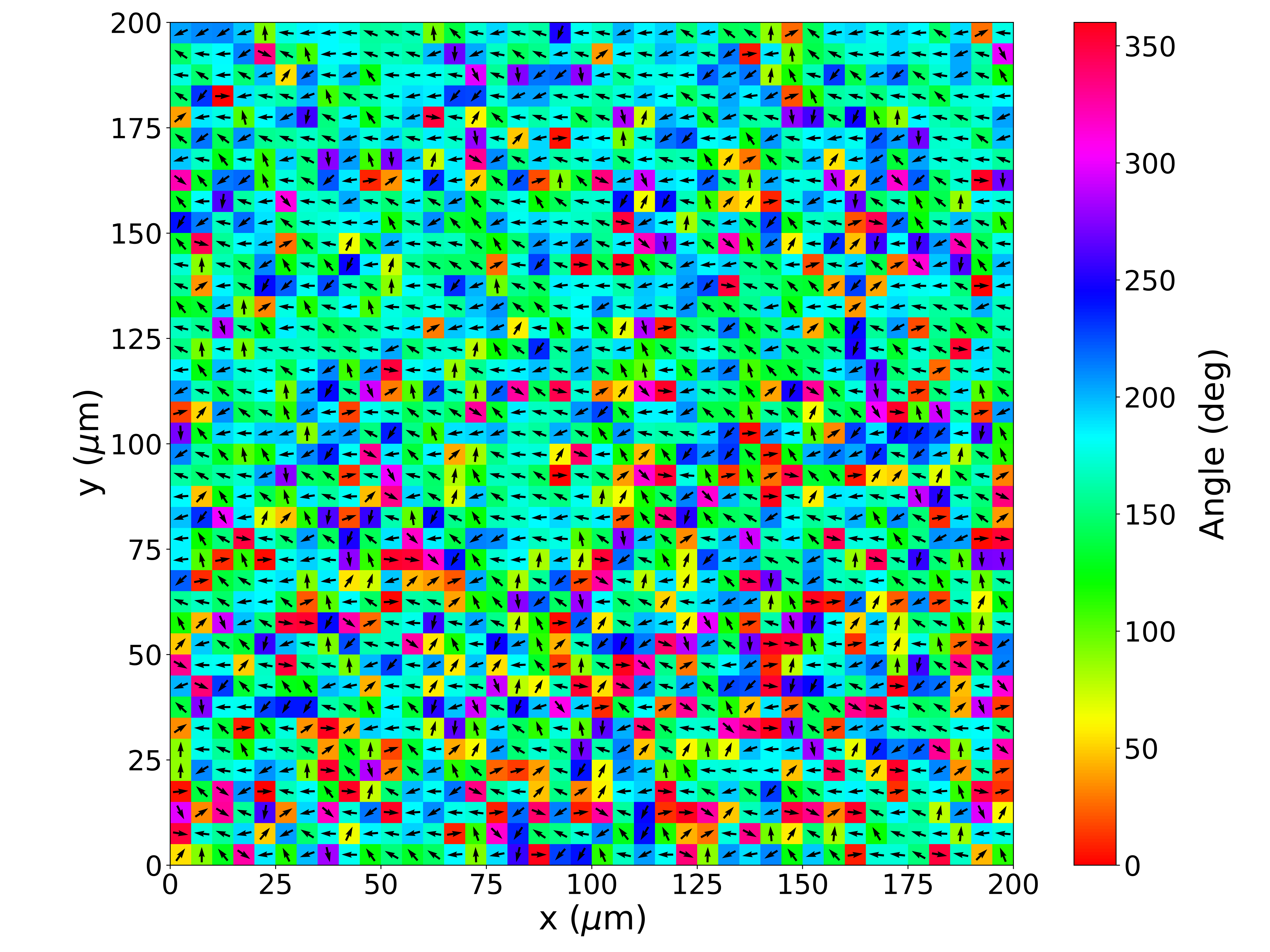}
\caption{Vector MOKE Map at 9 K (above the transition temperature).\label{fig:sup}}
\end{figure}